\documentstyle[preprint,aps]{revtex}

\begin{document}

\draft

\title{Novel Finite Temperature Conductivity in Quantum Hall Systems}
\author{Sudhansu S. Mandal, S. Ramaswamy, \cite{pa}
and V. Ravishankar}
\address{Department of Physics, Indian Institute of technology,
Kanpur -- 208 016, INDIA }

\maketitle

\begin{abstract}

We study quantum Hall systems (mainly the integer case) at finite
temperatures and show that there is a novel temperature dependence
even for a pure system, thanks to the `anomalous' nature of generators
of translation.
The deviation of Hall
conductivity from its zero temperature value is controlled
by a parameter $T_0 =\pi \rho /m^\ast N$ which is sample specific
and hence the universality of quantization is lost at finite
temperatures.

\end{abstract}



Quantum Hall (QH) systems have proved to be a rich source of
exploring many interesting and unexpected features of quantized
gauge theories.
The Hall conductivity which is quantized is topological
\cite{avron} in nature;
it appears (in the integer system that we mainly study here) as the
coefficient of Chern-Simons (CS) term which is induced by first order
quantum  corrections.
Further it is exact \cite{cole} -- with no higher order
corrections, and it does not suffer any renormalization.
It is also known that the Hall conductivity can be looked upon as a
manifestation of chiral anomaly \cite{jack,freid,ishik,hughes,widom},
inherited by the effectively planar system from
the parent three dimensional system.

Recall that in the Landau level problem at hand, the gauge
transformations get mixed up with the Euclidean transformations
in such a manner that
 the associated group is no more the (2+1) Euclidean
group $E_3$ , as one would normally have
for a pure system. Rather, it is
the group $M_3$ of magnetic translations which is a proper subgroup
of $E_3$. It is recognized that the transition from $E_3 \rightarrow
M_3$ is crucial. The generators of translation (or equivalently,
the operators for the centre of the orbit) do not commute \cite{chen}.
In his studies on the closely related CS superconductivity (CSS),
Fradkin \cite{frad} has designated this feature as `anomalous' and
has drawn detailed and explicit comparision with the well-known
Schwinger -- Anderson mechanism \cite{schwing,ander} which is a
proper field theoretic anomaly.

Here we do not attempt to rewrite the above mentioned non-commutativity
in the standard language of field theoretic anomaly.
However, we do believe in
the essential correctness of Fradkin's analogy, and  as an explicit
consequence we shall show that such an `anomaly' is responsible for a
novel temperature evolution of Hall conductivity $\sigma_H$
even for a {\it pure}
system. In this context we may recall that it is standard lore
\cite{prang} that
the presence of impurities, apart from its crucial role in stabilizing
the quantization (in form of plateaus) is further required to destroy
translational invariance in the system. It is believed that without
such a breaking QH effect (QHE) would be trivial.
Note that according to this argument, a uniform
distribution of impurities would still be insufficient
to give temperature dependence to Hall conductivity.
We show that the
Maxwell gauge interactions that are at play here belie such a naive
expectation.

Such a temperature dependence has been noticed in the allied albeit
rather academic example of CSS \cite{randj,ssman}. Even for QH
systems, Bellisard et. al. \cite{bellis},
have made rough estimates of the temperature
dependence of $\sigma_H$ for a pure system. We believe that this paper
presents, for the first time, a complete finite temperature (FT) analysis.
Further, we also hope that
the results obtained here will be verified experimentally.

Consider a system of (weakly) interacting electrons in two space
dimensions in the presence of a uniform external magnetic field of
strength $B$, confined to the direction perpendicular to the plane.
The strength is fine tuned such that $N$ Landau levels (LL)
are exactly filled.
In the presence of sufficiently high magnetic
field (as is relevant to our case),
the spins of the fermions would be `frozen' in the direction of
magnetic field. Therefore, one may treat the fermions as
spinless. The study of such a {\it spinless} system can be
accomplished with the
Lagrangian density,
\begin{equation}
{\cal L} = \psi^\ast iD_0\psi -\frac{1}{2m^\ast} \vert D_k \psi \vert^2
+\psi^\ast \mu \psi -eA_0^{\mbox{in}}\rho +\frac{1}{2}\int d^3x^\prime
A_0^{\mbox{in}} (x)V^{-1}(x-x^\prime)A_0^{\mbox{in}} (x^\prime) \, .
\label{eq1}
\end{equation}
Here $D_\nu =\partial_\nu -ie(A_\nu +A_0^{\mbox{in}}\delta_{\nu ,0} )$
(where $A_\nu $ is the external Maxwell gauge field and
$A_0^{\mbox{in}}$ is identified as internal scalar potential),
$\mu$ is the chemical potential, and $m^\ast$
and $\rho$ are the effective mass and the mean density of electrons
respectively. The fourth term in Eq.(\ref{eq1}) describes the charge
neutrality of the system. Finally, $V^{-1} (x-x^\prime)$ represents the
inverse of the instantaneous charge interaction potential
(in the operator
sense). The above Lagrangian density is equivalent to the usual
interaction term with quartic form of fermi fields,
which can be obtained by an integration of
$A_0^{\mbox{in}}$ field in Eq.~(\ref{eq1}).
Note also that the electrons interact with each other
via $1/r$ or some other short range potential,
i.e., the internal dynamics
is governed by the (3+1)-dimensional Maxwell Lagrangian
as is appropriate for
the medium.

The procedure for evaluating the FT
properties of the system with the
above Lagrangian density is standard. We do not discuss the details here
since they have been presented in the allied context of CSS
elegantly by
Randjbar-Daemi, Salam and Strathdee \cite{randj},
and has been extensively
used \cite{ssman}; in brief, we construct the partition function ($\beta
=1/T$ being the inverse temperature),
\begin{equation}
{\cal Z} = \int\, [dA_0^{\mbox{in}}][d\psi ][d\psi^\ast ] \exp \left[
-\int_0^\beta d\tau\,\int d^2r {\cal L}^{(E)} \right] \, ,
\label{eq2}
\end{equation}
which on integration over the fermionic fields, (by fixing of the saddle
point at the uniform background magnetic field $B$),
factors into ${\cal Z} ={\cal
Z}_B {\cal Z}_I$. Here ${\cal L}^{(E)}$ is the Euclidean version
of ${\cal L}$ in Eq.(\ref{eq1}). The background part of the partition
function is given by
$(1/A) \ln {\cal Z}_{\mbox{B}} = \rho_l \sum_{n=0}^\infty
\sum_{j=-\infty}^\infty \ln [\epsilon_n -\mu +i\omega_j ]$. Here
$\epsilon_n =(n+1/2)\omega_c$ (we have chosen the unit $\hbar =c =1$)
is the energy corresponding to $n$-th LL, where
$\omega_c = (e/m^\ast) B$ is  the
cyclotron frequency.
$\rho_l=m^\ast
\omega_c /2\pi $ is the degeneracy per unit area in each level,
and $A$ is
the area of the system.
Finally, $\omega_j =(2j+1)\pi/\beta $ is the fermionic
Matsubara frequency.
The corresponding thermodynamic potential is
obtained as $(\Omega /A)= -(\rho_l /\beta )\sum_{n=0}^\infty \ln
\left( 1+\exp
[-\beta (\epsilon_n -\mu )] \right)$,
from which all the properties for the system in
the background field can be
inferred.

Writing the partition function corresponding to the external probe as
${\cal Z}_I = \int [dA_0^{\mbox{in}}]\,\exp [-S_{\mbox{eff}}]$,
(where we have expanded the fermionic determinant upto
quadratic terms in powers of the fields
$A_0^{\mbox{in}}$ and the external
probe $A_\nu$
around the background field), we identify $S_{\mbox{eff}}$ with the
one-loop effective action which is obtained as
\begin{eqnarray}
S_{\mbox{eff}} &=& \frac{1}{2}\int d^3x\int d^3x^\prime
(A_\mu+A_\mu^{\mbox{in}}\delta_{\mu 0} ) \Pi^{\mu \nu} (x\, ,\,
x^\prime) (A_\nu+A_\nu^{\mbox{in}}\delta_{\nu 0} ) \nonumber \\
& &  -\frac{1}{2}\int d^3x
\int d^3 x^\prime A_0^{\mbox{in}}(x)V^{-1} (x-x^\prime)
A_0^{\mbox{in}} (x^\prime) \; .
\label{eq3}
\end{eqnarray}
The current correlation
functions $\Pi^{\mu \nu } (x,x^\prime)\equiv \delta
\langle j^\mu (x) \rangle /\delta {\cal A}_\nu (x^\prime)$,
where $j^\mu $ is the fermionic current,
and ${\cal A}_\nu $ is the sum of
all the gauge fields, have to be determined at the saddle point.
Using Galilean and
gauge invariance, we write (in the momentum space)
\begin{eqnarray}
\Pi^{\mu \nu} (\omega \, ,\, {\bf q}) &=& \Pi_0 (\omega\, ,\, {\bf q})
(q^2g^{\mu \nu} -q^\mu q^\nu )+(\Pi_2-\Pi_0) (\omega \, ,\, {\bf q})
\nonumber  \\
& & \times ({\bf q}^2 \delta^{ij}-q^iq^j)\delta^{\mu i}\delta^{\nu j}
+i\Pi_1 (\omega \, ,\, {\bf q}) \epsilon^{\mu \nu \lambda}q_\lambda \; ,
\label{eq4}
\end{eqnarray}
At FT, $\Pi_0$ acquires a pole at ${\bf q}^2 =0$, i.e.,
$\Pi_0=\bar{\Pi}_0+\Gamma/{\bf q}^2 $.
Note that this pole exists in the limit
$\omega =0 \, ,\, {\bf q}^2 \rightarrow 0$.
On the other hand, for ${\bf q}^2 = 0\, ,\, \omega \rightarrow 0 $,
$\Gamma \equiv 0$ (no pole exists).
However, the limits do commute as far as other form factors are
concerned.
The FT responses of the system are driven by the temperature
behaviour of these form factors.

As we are interested in the low energy response of the system, it is
sufficient that we evaluate the form factors at $\omega =0\, ,\,
{\bf q}^2 =0$ (keeping in mind the above mentioned singularity
of $\Pi_0$). Therefore we obtain
\begin{mathletters}
\label{eq5}
\begin{eqnarray}
\bar{\Pi}_0(0,0) &=& \frac{e^2}{4\pi}\sum_{n=0}^\infty \sum_{m=0}^\infty
\frac{(f_m-f_n)}{\epsilon_n -\epsilon_m}\left[ (n+1)\delta_{m,n+1}
+n\delta_{m,n-1} -(2n+1)\delta_{m,n} \right]  \nonumber \\
& = & \frac{e^2}{2\pi\omega_c}\sum_{n=0}^\infty
f_n -\frac{e^2}{4\pi} \beta
\sum_{n=0}^\infty (2n+1) f_n (1-f_n) \, ,  \\
\Pi_1 (0,0) &=& \bar{\Pi}_0 \omega_c \, , \\
\Pi_2 (0,0) &=& \frac{e^2\omega_c}{8\pi m^\ast}
 \sum_{m=0}^\infty \sum_{n=0}^\infty
 \frac{f_m -f_n}{\epsilon_n -\epsilon_m}
 \left[ 3(n+1)^2\delta_{m,\, n+1} +3n^2\delta_{m,\, n-1} \right.
  \nonumber  \\
 & & \left. -(n+1)(n+2)\delta_{m,\, n+2} -n(n-1)\delta_{m,\, n-2}
 -(2n+1)^2 \delta_{m,\, n} \right]    \nonumber \\
 &=& \frac{e^2}{2\pi m^\ast}\sum_{n=0}^\infty
 (2n+1)f_n -\frac{e^2}{8\pi m^\ast
  } \beta\omega_c \sum_{n=0}^\infty (2n+1)^2 f_n (1-f_n)  \\
\Gamma &=& \frac{e^2 m^\ast}{2\pi}\omega_c \sum_{n=0}^\infty
\sum_{m=0}^\infty \frac{f_m -f_n}{\epsilon_n -\epsilon_m} \delta_{mn}
\nonumber  \\
&=& \frac{e^2 m^\ast}{2\pi}\beta \omega_c \sum_{n=0}^\infty
f_n (1-f_n) \, ,
\end{eqnarray}
\end{mathletters}
with $f_n=[1+\exp (\beta [\epsilon_n -\mu ])]^{-1}$.

For each of the form factors (except $\Gamma$) the first term in
Eq.~(\ref{eq5}) contributes for all values of $\omega $
and is continuous at
$\omega =0$. On the other hand, the second term in the expressions
for $\bar{\Pi}_0$, $\Pi_1$ and $\Pi_2$, as well as $\Gamma$, has a
singular dependence on $\omega $, i.e., it is non-vanishing {\it only}
at $\omega =0$. Indeed,
the former non-singular contribution is caused by
the inter LL transitions,
while the latter singular ones owe their existence
to the intra LL transitions in the virtual process shown in Fig.~1.
It is significant that the entire temperature dependence of $\Pi_1$,
and hence Hall conductivity $\sigma_H$ as obtained below, is singular.
We observe that this is a direct consequence of the `anomaly' in
the translation generators at hand; for, but for the infinite
degeneracy in the LL, this novel temperature dependence in $\Pi_1$
would not survive in the thermodynamic limit.

The parity and time reversal violating form factor $\Pi_1$ which is the
coefficient of CS term in the effective action has interesting
properties. It is purely topological. Moreover, it does not get
renormalized by the higher order calculation of correlation function
according to Coleman-Hill theorem \cite{cole}. Therefore, $\Pi_1$
in Eq.~(\ref{eq5}) is exact.

A straight forward linear response analysis from
Eqs.~(\ref{eq3}--\ref{eq5})
yields the Hall conductivity to be
\begin{equation}
\sigma_H = \Pi_1 (0,0) = \frac{e^2}{2\pi}\sum_{n=0}^\infty f_n -
\frac{e^2}{4\pi}\beta\omega_c \sum_{n=0}^\infty (2n+1)f_n (1-f_n)\, ,
\label{eq6}
\end{equation}
subject to the condition $\lim_{{\bf q}^2 \rightarrow 0} V(|{\bf q}|)
{\bf q}^2 =0$. In other words, for any short ranged potential, Hall
conductivity is exactly the parity and time reversal violating form
factor. Note that the electron-electron interaction does not
otherwise play any major role in this case.
Therefore the emergence of $\sigma_H$
is also purely topological and exact.
Note that the diagonal conductivity vanishes by
virtue of the purity of the system.

At $T=0$, $\sigma_H$ is quantized to the value $\nu(e^2/2\pi)$,
where the filling fraction $\nu =N$ (an integer).
The quantization is `universal', i.e.,
it does not depend on the microscopic
details of the system. Since QHE has been observed at very low
temperatures, a low temperature expansion of $\sigma_H$ should suffice.
In that case, it is analytically evaluated as a perturbation in $\exp
[-\beta\omega_c /2 ]$ (see Refs. 13 and 14 for details of calculation)
and is found to be
\begin{equation}
\sigma_H (T)=\frac{e^2}{2\pi}N(1-4y) \, ,
\label{eq7}
\end{equation}
where $y=(T_0/T) \exp [-T_0/T]$, with $T_0=\pi\rho/m^\ast N$.
Bellissard et al \cite{bellis} have also discussed the temperature
dependent Hall conductivity using Kubo formula for the sample of
infinite relaxation time. To compare their results with our exact
result, we note that they only
make an approximate estimation of the change in $\sigma_H$ where
they get the correct exponential term, but miss the crucial
prefactor multiplying it.
 From the expression (\ref{eq7}), it is clear that the novel
temperature dependence is indeed accompanied by corresponding deviation
from universality of quantization in virtue of its dependence on the
parameter $T_0$ which is the only sample specific parameter that enters
the analysis. In fact, at any temperature, although we cannot evaluate
$\sigma_H (T)$ analytically, it is easy to check that the Hall defect
\begin{equation}
{\cal R} \equiv \left\vert \frac{\sigma_H(T)-\sigma_H (0)}{\sigma_H
(0)} \right\vert
\label{eq8}
\end{equation}
is a function of the dimensionless variable $T_0/T$. This type of
temperature dependence and the specific form of $T_0$ is a
reflection of the
fundamental energy scale $\omega_c$.

If we fix the value of ${\cal R}$,
the temperature $T_{{\cal R}}$ at which
the defect ${\cal R}$ would occur for QHE follows a simple
expression
\begin{equation}
\frac{T_0}{T_{{\cal R}}} =C \, ,
\label{eq9}
\end{equation}
where the constant $C$ is independent of the sample.
For example, the value of $C$ at ${\cal R} = 10^{-n}$ is approximately
given by $0.325+0.776(n+1)$ for $3\leq n\leq 8$. Fig.~2 shows how
$T_{{\cal R}}$ depends on ${\cal R}$ for $\nu =1$ and $2$ over a
range 0.01 ppm to $0.1\%$ for a specific choice of $\rho /m^\ast$.

We have considered integer QHE here for simplicity's sake.
We report that a similar analysis holds for fractional QHE
within the composite fermion model \cite{jain}. The
analysis in this case involves other aspects such as the mean field
ansatz, which we shall defer to a different paper.

Before we conclude,
we observe that there is already a wealth of experimental
infomation available on FT QHE for both integer and
fractional case \cite{yoshi,cage,chang} showing the deviation from
quantization at the central value of $B$. We therefore believe that
it is not impossible to verify experimentally the effect predicted
here.
Admittedly the contribution from impurity dominates
over the one at hand \cite{pub}.
However we hope that $\sigma_H (T)$ will
be measured by varying the disorder (keeping
other parameters fixed).
( See \cite{expt} for one such
experiment). One may then extrapolate the result to zero impurity
concentration, or, if the impurity contribution is well understood,
merely subtract that part to extract the required temperature
dependence.

Acknowledgements:
We thank S. D. Joglekar and J. K. Bhattacharjee
    for helpful discussions.

\newpage

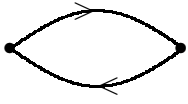
\begin{figure}
\setlength{\unitlength}{0.240900pt}
\ifx\plotpoint\undefined\newsavebox{\plotpoint}\fi
\sbox{\plotpoint}{\rule[-0.175pt]{0.350pt}{0.350pt}}%
\begin{picture}(600,390)(0,0)
\tenrm
\sbox{\plotpoint}{\rule[-0.350pt]{0.700pt}{0.700pt}}%
\put(400,277){\makebox(0,0)[r]{$ > $}}
\put(400,158){\makebox(0,0)[l]{$ < $}}
\put(264,218){\circle*{15}}
\put(533,218){\circle*{15}}
\put(264,218){\usebox{\plotpoint}}
\put(264,218){\rule[-0.350pt]{0.723pt}{0.700pt}}
\put(267,219){\usebox{\plotpoint}}
\put(268,220){\usebox{\plotpoint}}
\put(269,221){\usebox{\plotpoint}}
\put(270,222){\usebox{\plotpoint}}
\put(272,223){\usebox{\plotpoint}}
\put(273,224){\usebox{\plotpoint}}
\put(275,225){\usebox{\plotpoint}}
\put(276,226){\usebox{\plotpoint}}
\put(278,227){\usebox{\plotpoint}}
\put(279,228){\usebox{\plotpoint}}
\put(280,229){\usebox{\plotpoint}}
\put(281,230){\usebox{\plotpoint}}
\put(283,231){\rule[-0.350pt]{0.723pt}{0.700pt}}
\put(286,232){\usebox{\plotpoint}}
\put(287,233){\usebox{\plotpoint}}
\put(289,234){\usebox{\plotpoint}}
\put(290,235){\usebox{\plotpoint}}
\put(291,236){\usebox{\plotpoint}}
\put(292,237){\usebox{\plotpoint}}
\put(294,238){\usebox{\plotpoint}}
\put(295,239){\usebox{\plotpoint}}
\put(297,240){\rule[-0.350pt]{0.723pt}{0.700pt}}
\put(300,241){\usebox{\plotpoint}}
\put(301,242){\usebox{\plotpoint}}
\put(302,243){\usebox{\plotpoint}}
\put(303,244){\usebox{\plotpoint}}
\put(305,245){\rule[-0.350pt]{0.723pt}{0.700pt}}
\put(308,246){\usebox{\plotpoint}}
\put(309,247){\usebox{\plotpoint}}
\put(311,248){\usebox{\plotpoint}}
\put(312,249){\usebox{\plotpoint}}
\put(313,250){\rule[-0.350pt]{0.723pt}{0.700pt}}
\put(316,251){\usebox{\plotpoint}}
\put(317,252){\usebox{\plotpoint}}
\put(319,253){\rule[-0.350pt]{0.723pt}{0.700pt}}
\put(322,254){\usebox{\plotpoint}}
\put(323,255){\usebox{\plotpoint}}
\put(324,256){\rule[-0.350pt]{0.723pt}{0.700pt}}
\put(327,257){\usebox{\plotpoint}}
\put(328,258){\usebox{\plotpoint}}
\put(330,259){\rule[-0.350pt]{0.723pt}{0.700pt}}
\put(333,260){\usebox{\plotpoint}}
\put(335,261){\rule[-0.350pt]{0.723pt}{0.700pt}}
\put(338,262){\usebox{\plotpoint}}
\put(339,263){\usebox{\plotpoint}}
\put(341,264){\rule[-0.350pt]{0.723pt}{0.700pt}}
\put(344,265){\usebox{\plotpoint}}
\put(346,266){\rule[-0.350pt]{0.723pt}{0.700pt}}
\put(349,267){\rule[-0.350pt]{0.723pt}{0.700pt}}
\put(352,268){\rule[-0.350pt]{0.723pt}{0.700pt}}
\put(355,269){\usebox{\plotpoint}}
\put(357,270){\rule[-0.350pt]{0.723pt}{0.700pt}}
\put(360,271){\rule[-0.350pt]{0.723pt}{0.700pt}}
\put(363,272){\rule[-0.350pt]{1.204pt}{0.700pt}}
\put(368,273){\rule[-0.350pt]{0.723pt}{0.700pt}}
\put(371,274){\rule[-0.350pt]{1.445pt}{0.700pt}}
\put(377,275){\rule[-0.350pt]{1.204pt}{0.700pt}}
\put(382,276){\rule[-0.350pt]{1.927pt}{0.700pt}}
\put(390,277){\rule[-0.350pt]{5.300pt}{0.700pt}}
\put(412,276){\rule[-0.350pt]{2.168pt}{0.700pt}}
\put(421,275){\rule[-0.350pt]{1.204pt}{0.700pt}}
\put(426,274){\rule[-0.350pt]{1.445pt}{0.700pt}}
\put(432,273){\usebox{\plotpoint}}
\put(434,272){\rule[-0.350pt]{1.445pt}{0.700pt}}
\put(440,271){\rule[-0.350pt]{0.723pt}{0.700pt}}
\put(443,270){\usebox{\plotpoint}}
\put(445,269){\rule[-0.350pt]{0.723pt}{0.700pt}}
\put(448,268){\rule[-0.350pt]{0.723pt}{0.700pt}}
\put(451,267){\rule[-0.350pt]{0.723pt}{0.700pt}}
\put(454,266){\usebox{\plotpoint}}
\put(456,265){\rule[-0.350pt]{0.723pt}{0.700pt}}
\put(459,264){\usebox{\plotpoint}}
\put(460,263){\usebox{\plotpoint}}
\put(462,262){\rule[-0.350pt]{0.723pt}{0.700pt}}
\put(465,261){\usebox{\plotpoint}}
\put(467,260){\rule[-0.350pt]{0.723pt}{0.700pt}}
\put(470,259){\usebox{\plotpoint}}
\put(471,258){\usebox{\plotpoint}}
\put(473,257){\rule[-0.350pt]{0.723pt}{0.700pt}}
\put(476,256){\usebox{\plotpoint}}
\put(477,255){\usebox{\plotpoint}}
\put(478,254){\rule[-0.350pt]{0.723pt}{0.700pt}}
\put(481,253){\usebox{\plotpoint}}
\put(482,252){\usebox{\plotpoint}}
\put(484,251){\rule[-0.350pt]{0.723pt}{0.700pt}}
\put(487,250){\usebox{\plotpoint}}
\put(488,249){\usebox{\plotpoint}}
\put(489,248){\usebox{\plotpoint}}
\put(490,247){\usebox{\plotpoint}}
\put(492,246){\rule[-0.350pt]{0.723pt}{0.700pt}}
\put(495,245){\usebox{\plotpoint}}
\put(496,244){\usebox{\plotpoint}}
\put(498,243){\usebox{\plotpoint}}
\put(499,242){\usebox{\plotpoint}}
\put(500,241){\rule[-0.350pt]{0.723pt}{0.700pt}}
\put(503,240){\usebox{\plotpoint}}
\put(504,239){\usebox{\plotpoint}}
\put(506,238){\usebox{\plotpoint}}
\put(507,237){\usebox{\plotpoint}}
\put(509,236){\usebox{\plotpoint}}
\put(510,235){\usebox{\plotpoint}}
\put(511,234){\usebox{\plotpoint}}
\put(512,233){\usebox{\plotpoint}}
\put(514,232){\rule[-0.350pt]{0.723pt}{0.700pt}}
\put(517,231){\usebox{\plotpoint}}
\put(518,230){\usebox{\plotpoint}}
\put(520,229){\usebox{\plotpoint}}
\put(521,228){\usebox{\plotpoint}}
\put(522,227){\usebox{\plotpoint}}
\put(523,226){\usebox{\plotpoint}}
\put(525,225){\usebox{\plotpoint}}
\put(526,224){\usebox{\plotpoint}}
\put(528,223){\usebox{\plotpoint}}
\put(529,222){\usebox{\plotpoint}}
\put(531,221){\usebox{\plotpoint}}
\put(532,220){\usebox{\plotpoint}}
\put(533,219){\rule[-0.350pt]{0.723pt}{0.700pt}}
\put(264,218){\usebox{\plotpoint}}
\put(264,218){\usebox{\plotpoint}}
\put(265,217){\usebox{\plotpoint}}
\put(267,216){\usebox{\plotpoint}}
\put(268,215){\usebox{\plotpoint}}
\put(269,214){\usebox{\plotpoint}}
\put(270,213){\usebox{\plotpoint}}
\put(272,212){\usebox{\plotpoint}}
\put(273,211){\usebox{\plotpoint}}
\put(275,210){\usebox{\plotpoint}}
\put(276,209){\usebox{\plotpoint}}
\put(278,208){\usebox{\plotpoint}}
\put(279,207){\usebox{\plotpoint}}
\put(280,206){\usebox{\plotpoint}}
\put(281,205){\usebox{\plotpoint}}
\put(283,204){\rule[-0.350pt]{0.723pt}{0.700pt}}
\put(286,203){\usebox{\plotpoint}}
\put(287,202){\usebox{\plotpoint}}
\put(289,201){\usebox{\plotpoint}}
\put(290,200){\usebox{\plotpoint}}
\put(291,199){\usebox{\plotpoint}}
\put(292,198){\usebox{\plotpoint}}
\put(294,197){\usebox{\plotpoint}}
\put(295,196){\usebox{\plotpoint}}
\put(297,195){\rule[-0.350pt]{0.723pt}{0.700pt}}
\put(300,194){\usebox{\plotpoint}}
\put(301,193){\usebox{\plotpoint}}
\put(302,192){\usebox{\plotpoint}}
\put(303,191){\usebox{\plotpoint}}
\put(305,190){\rule[-0.350pt]{0.723pt}{0.700pt}}
\put(308,189){\usebox{\plotpoint}}
\put(309,188){\usebox{\plotpoint}}
\put(311,187){\usebox{\plotpoint}}
\put(312,186){\usebox{\plotpoint}}
\put(313,185){\rule[-0.350pt]{0.723pt}{0.700pt}}
\put(316,184){\usebox{\plotpoint}}
\put(317,183){\usebox{\plotpoint}}
\put(319,182){\rule[-0.350pt]{0.723pt}{0.700pt}}
\put(322,181){\usebox{\plotpoint}}
\put(323,180){\usebox{\plotpoint}}
\put(324,179){\rule[-0.350pt]{0.723pt}{0.700pt}}
\put(327,178){\usebox{\plotpoint}}
\put(328,177){\usebox{\plotpoint}}
\put(330,176){\rule[-0.350pt]{0.723pt}{0.700pt}}
\put(333,175){\usebox{\plotpoint}}
\put(335,174){\rule[-0.350pt]{0.723pt}{0.700pt}}
\put(338,173){\usebox{\plotpoint}}
\put(339,172){\usebox{\plotpoint}}
\put(341,171){\rule[-0.350pt]{0.723pt}{0.700pt}}
\put(344,170){\usebox{\plotpoint}}
\put(346,169){\rule[-0.350pt]{0.723pt}{0.700pt}}
\put(349,168){\rule[-0.350pt]{0.723pt}{0.700pt}}
\put(352,167){\rule[-0.350pt]{0.723pt}{0.700pt}}
\put(355,166){\usebox{\plotpoint}}
\put(357,165){\rule[-0.350pt]{0.723pt}{0.700pt}}
\put(360,164){\rule[-0.350pt]{0.723pt}{0.700pt}}
\put(363,163){\rule[-0.350pt]{1.204pt}{0.700pt}}
\put(368,162){\rule[-0.350pt]{0.723pt}{0.700pt}}
\put(371,161){\rule[-0.350pt]{1.445pt}{0.700pt}}
\put(377,160){\rule[-0.350pt]{1.204pt}{0.700pt}}
\put(382,159){\rule[-0.350pt]{1.927pt}{0.700pt}}
\put(390,158){\rule[-0.350pt]{5.300pt}{0.700pt}}
\put(412,159){\rule[-0.350pt]{2.168pt}{0.700pt}}
\put(421,160){\rule[-0.350pt]{1.204pt}{0.700pt}}
\put(426,161){\rule[-0.350pt]{1.445pt}{0.700pt}}
\put(432,162){\usebox{\plotpoint}}
\put(434,163){\rule[-0.350pt]{1.445pt}{0.700pt}}
\put(440,164){\rule[-0.350pt]{0.723pt}{0.700pt}}
\put(443,165){\usebox{\plotpoint}}
\put(445,166){\rule[-0.350pt]{0.723pt}{0.700pt}}
\put(448,167){\rule[-0.350pt]{0.723pt}{0.700pt}}
\put(451,168){\rule[-0.350pt]{0.723pt}{0.700pt}}
\put(454,169){\usebox{\plotpoint}}
\put(456,170){\rule[-0.350pt]{0.723pt}{0.700pt}}
\put(459,171){\usebox{\plotpoint}}
\put(460,172){\usebox{\plotpoint}}
\put(462,173){\rule[-0.350pt]{0.723pt}{0.700pt}}
\put(465,174){\usebox{\plotpoint}}
\put(467,175){\rule[-0.350pt]{0.723pt}{0.700pt}}
\put(470,176){\usebox{\plotpoint}}
\put(471,177){\usebox{\plotpoint}}
\put(473,178){\rule[-0.350pt]{0.723pt}{0.700pt}}
\put(476,179){\usebox{\plotpoint}}
\put(477,180){\usebox{\plotpoint}}
\put(478,181){\rule[-0.350pt]{0.723pt}{0.700pt}}
\put(481,182){\usebox{\plotpoint}}
\put(482,183){\usebox{\plotpoint}}
\put(484,184){\rule[-0.350pt]{0.723pt}{0.700pt}}
\put(487,185){\usebox{\plotpoint}}
\put(488,186){\usebox{\plotpoint}}
\put(489,187){\usebox{\plotpoint}}
\put(490,188){\usebox{\plotpoint}}
\put(492,189){\rule[-0.350pt]{0.723pt}{0.700pt}}
\put(495,190){\usebox{\plotpoint}}
\put(496,191){\usebox{\plotpoint}}
\put(498,192){\usebox{\plotpoint}}
\put(499,193){\usebox{\plotpoint}}
\put(500,194){\rule[-0.350pt]{0.723pt}{0.700pt}}
\put(503,195){\usebox{\plotpoint}}
\put(504,196){\usebox{\plotpoint}}
\put(506,197){\usebox{\plotpoint}}
\put(507,198){\usebox{\plotpoint}}
\put(509,199){\usebox{\plotpoint}}
\put(510,200){\usebox{\plotpoint}}
\put(511,201){\usebox{\plotpoint}}
\put(512,202){\usebox{\plotpoint}}
\put(514,203){\rule[-0.350pt]{0.723pt}{0.700pt}}
\put(517,204){\usebox{\plotpoint}}
\put(518,205){\usebox{\plotpoint}}
\put(520,206){\usebox{\plotpoint}}
\put(521,207){\usebox{\plotpoint}}
\put(522,208){\usebox{\plotpoint}}
\put(523,209){\usebox{\plotpoint}}
\put(525,210){\usebox{\plotpoint}}
\put(526,211){\usebox{\plotpoint}}
\put(528,212){\usebox{\plotpoint}}
\put(529,213){\usebox{\plotpoint}}
\put(531,214){\usebox{\plotpoint}}
\put(532,215){\usebox{\plotpoint}}
\put(533,216){\rule[-0.350pt]{0.723pt}{0.700pt}}
\end{picture}
\caption{Vacuum polarization diagram for the virtual process.}
\end{figure}


\begin{figure}
\setlength{\unitlength}{0.240900pt}
\ifx\plotpoint\undefined\newsavebox{\plotpoint}\fi
\sbox{\plotpoint}{\rule[-0.175pt]{0.350pt}{0.350pt}}%
\begin{picture}(1005,900)(0,0)
\tenrm
\put(264,158){\rule[-0.175pt]{4.818pt}{0.350pt}}
\put(242,158){\makebox(0,0)[r]{$0.2$}}
\put(921,158){\rule[-0.175pt]{4.818pt}{0.350pt}}
\put(264,263){\rule[-0.175pt]{4.818pt}{0.350pt}}
\put(242,263){\makebox(0,0)[r]{$0.4$}}
\put(921,263){\rule[-0.175pt]{4.818pt}{0.350pt}}
\put(264,368){\rule[-0.175pt]{4.818pt}{0.350pt}}
\put(242,368){\makebox(0,0)[r]{$0.6$}}
\put(921,368){\rule[-0.175pt]{4.818pt}{0.350pt}}
\put(264,473){\rule[-0.175pt]{4.818pt}{0.350pt}}
\put(242,473){\makebox(0,0)[r]{$0.8$}}
\put(921,473){\rule[-0.175pt]{4.818pt}{0.350pt}}
\put(264,577){\rule[-0.175pt]{4.818pt}{0.350pt}}
\put(242,577){\makebox(0,0)[r]{$1.0$}}
\put(921,577){\rule[-0.175pt]{4.818pt}{0.350pt}}
\put(264,682){\rule[-0.175pt]{4.818pt}{0.350pt}}
\put(242,682){\makebox(0,0)[r]{$1.2$}}
\put(921,682){\rule[-0.175pt]{4.818pt}{0.350pt}}
\put(264,787){\rule[-0.175pt]{4.818pt}{0.350pt}}
\put(242,787){\makebox(0,0)[r]{$1.4$}}
\put(921,787){\rule[-0.175pt]{4.818pt}{0.350pt}}
\put(264,158){\rule[-0.175pt]{0.350pt}{4.818pt}}
\put(264,113){\makebox(0,0){$10^{-8}$}}
\put(264,767){\rule[-0.175pt]{0.350pt}{4.818pt}}
\put(346,158){\rule[-0.175pt]{0.350pt}{2.409pt}}
\put(346,777){\rule[-0.175pt]{0.350pt}{2.409pt}}
\put(378,158){\rule[-0.175pt]{0.350pt}{2.409pt}}
\put(378,777){\rule[-0.175pt]{0.350pt}{2.409pt}}
\put(399,158){\rule[-0.175pt]{0.350pt}{4.818pt}}
\put(399,113){\makebox(0,0){$10^{-7}$}}
\put(399,767){\rule[-0.175pt]{0.350pt}{4.818pt}}
\put(481,158){\rule[-0.175pt]{0.350pt}{2.409pt}}
\put(481,777){\rule[-0.175pt]{0.350pt}{2.409pt}}
\put(514,158){\rule[-0.175pt]{0.350pt}{2.409pt}}
\put(514,777){\rule[-0.175pt]{0.350pt}{2.409pt}}
\put(535,158){\rule[-0.175pt]{0.350pt}{4.818pt}}
\put(535,113){\makebox(0,0){$10^{-6}$}}
\put(535,767){\rule[-0.175pt]{0.350pt}{4.818pt}}
\put(616,158){\rule[-0.175pt]{0.350pt}{2.409pt}}
\put(616,777){\rule[-0.175pt]{0.350pt}{2.409pt}}
\put(649,158){\rule[-0.175pt]{0.350pt}{2.409pt}}
\put(649,777){\rule[-0.175pt]{0.350pt}{2.409pt}}
\put(670,158){\rule[-0.175pt]{0.350pt}{4.818pt}}
\put(670,113){\makebox(0,0){$10^{-5}$}}
\put(670,767){\rule[-0.175pt]{0.350pt}{4.818pt}}
\put(752,158){\rule[-0.175pt]{0.350pt}{2.409pt}}
\put(752,777){\rule[-0.175pt]{0.350pt}{2.409pt}}
\put(785,158){\rule[-0.175pt]{0.350pt}{2.409pt}}
\put(785,777){\rule[-0.175pt]{0.350pt}{2.409pt}}
\put(806,158){\rule[-0.175pt]{0.350pt}{4.818pt}}
\put(806,113){\makebox(0,0){$10^{-4}$}}
\put(806,767){\rule[-0.175pt]{0.350pt}{4.818pt}}
\put(887,158){\rule[-0.175pt]{0.350pt}{2.409pt}}
\put(887,777){\rule[-0.175pt]{0.350pt}{2.409pt}}
\put(920,158){\rule[-0.175pt]{0.350pt}{2.409pt}}
\put(920,777){\rule[-0.175pt]{0.350pt}{2.409pt}}
\put(941,158){\rule[-0.175pt]{0.350pt}{4.818pt}}
\put(941,113){\makebox(0,0){$10^{-3}$}}
\put(941,767){\rule[-0.175pt]{0.350pt}{4.818pt}}
\put(264,158){\rule[-0.175pt]{163.089pt}{0.350pt}}
\put(941,158){\rule[-0.175pt]{0.350pt}{151.526pt}}
\put(264,787){\rule[-0.175pt]{163.089pt}{0.350pt}}
\put(45,472){\makebox(0,0)[l]{\shortstack{$T_{\cal R}$(K)}}}
\put(602,68){\makebox(0,0){${\cal R}$}}
\put(806,672){\makebox(0,0)[l]{a}}
\put(806,316){\makebox(0,0)[l]{b}}
\put(264,158){\rule[-0.175pt]{0.350pt}{151.526pt}}
\sbox{\plotpoint}{\rule[-0.350pt]{0.700pt}{0.700pt}}%
\put(264,382){\rule[-0.350pt]{0.898pt}{0.700pt}}
\put(267,383){\rule[-0.350pt]{0.898pt}{0.700pt}}
\put(271,384){\rule[-0.350pt]{0.898pt}{0.700pt}}
\put(275,385){\rule[-0.350pt]{0.898pt}{0.700pt}}
\put(278,386){\rule[-0.350pt]{0.898pt}{0.700pt}}
\put(282,387){\rule[-0.350pt]{0.898pt}{0.700pt}}
\put(286,388){\rule[-0.350pt]{0.898pt}{0.700pt}}
\put(290,389){\rule[-0.350pt]{0.898pt}{0.700pt}}
\put(293,390){\rule[-0.350pt]{0.898pt}{0.700pt}}
\put(297,391){\rule[-0.350pt]{0.898pt}{0.700pt}}
\put(301,392){\rule[-0.350pt]{0.898pt}{0.700pt}}
\put(305,393){\rule[-0.350pt]{0.964pt}{0.700pt}}
\put(309,394){\rule[-0.350pt]{0.964pt}{0.700pt}}
\put(313,395){\rule[-0.350pt]{0.964pt}{0.700pt}}
\put(317,396){\rule[-0.350pt]{0.964pt}{0.700pt}}
\put(321,397){\rule[-0.350pt]{0.964pt}{0.700pt}}
\put(325,398){\rule[-0.350pt]{0.964pt}{0.700pt}}
\put(329,399){\rule[-0.350pt]{0.819pt}{0.700pt}}
\put(332,400){\rule[-0.350pt]{0.819pt}{0.700pt}}
\put(335,401){\rule[-0.350pt]{0.819pt}{0.700pt}}
\put(339,402){\rule[-0.350pt]{0.819pt}{0.700pt}}
\put(342,403){\rule[-0.350pt]{0.819pt}{0.700pt}}
\put(346,404){\rule[-0.350pt]{0.783pt}{0.700pt}}
\put(349,405){\rule[-0.350pt]{0.783pt}{0.700pt}}
\put(352,406){\rule[-0.350pt]{0.783pt}{0.700pt}}
\put(355,407){\rule[-0.350pt]{0.783pt}{0.700pt}}
\put(359,408){\rule[-0.350pt]{0.803pt}{0.700pt}}
\put(362,409){\rule[-0.350pt]{0.803pt}{0.700pt}}
\put(365,410){\rule[-0.350pt]{0.803pt}{0.700pt}}
\put(369,411){\rule[-0.350pt]{0.723pt}{0.700pt}}
\put(372,412){\rule[-0.350pt]{0.723pt}{0.700pt}}
\put(375,413){\rule[-0.350pt]{0.723pt}{0.700pt}}
\put(378,414){\rule[-0.350pt]{0.964pt}{0.700pt}}
\put(382,415){\rule[-0.350pt]{0.964pt}{0.700pt}}
\put(386,416){\usebox{\plotpoint}}
\put(388,417){\usebox{\plotpoint}}
\put(390,418){\usebox{\plotpoint}}
\put(393,419){\rule[-0.350pt]{0.723pt}{0.700pt}}
\put(396,420){\rule[-0.350pt]{0.723pt}{0.700pt}}
\put(399,421){\rule[-0.350pt]{0.826pt}{0.700pt}}
\put(402,422){\rule[-0.350pt]{0.826pt}{0.700pt}}
\put(405,423){\rule[-0.350pt]{0.826pt}{0.700pt}}
\put(409,424){\rule[-0.350pt]{0.826pt}{0.700pt}}
\put(412,425){\rule[-0.350pt]{0.826pt}{0.700pt}}
\put(416,426){\rule[-0.350pt]{0.826pt}{0.700pt}}
\put(419,427){\rule[-0.350pt]{0.826pt}{0.700pt}}
\put(423,428){\usebox{\plotpoint}}
\put(425,429){\usebox{\plotpoint}}
\put(428,430){\usebox{\plotpoint}}
\put(431,431){\usebox{\plotpoint}}
\put(434,432){\usebox{\plotpoint}}
\put(437,433){\usebox{\plotpoint}}
\put(440,434){\usebox{\plotpoint}}
\put(442,435){\usebox{\plotpoint}}
\put(445,436){\usebox{\plotpoint}}
\put(447,437){\usebox{\plotpoint}}
\put(450,438){\usebox{\plotpoint}}
\put(453,439){\usebox{\plotpoint}}
\put(455,440){\usebox{\plotpoint}}
\put(458,441){\usebox{\plotpoint}}
\put(461,442){\usebox{\plotpoint}}
\put(464,443){\rule[-0.350pt]{0.723pt}{0.700pt}}
\put(467,444){\rule[-0.350pt]{0.723pt}{0.700pt}}
\put(470,445){\rule[-0.350pt]{0.723pt}{0.700pt}}
\put(473,446){\usebox{\plotpoint}}
\put(475,447){\usebox{\plotpoint}}
\put(478,448){\usebox{\plotpoint}}
\put(481,449){\rule[-0.350pt]{0.843pt}{0.700pt}}
\put(484,450){\rule[-0.350pt]{0.843pt}{0.700pt}}
\put(488,451){\usebox{\plotpoint}}
\put(490,452){\usebox{\plotpoint}}
\put(492,453){\usebox{\plotpoint}}
\put(494,454){\rule[-0.350pt]{0.723pt}{0.700pt}}
\put(497,455){\rule[-0.350pt]{0.723pt}{0.700pt}}
\put(500,456){\usebox{\plotpoint}}
\put(502,457){\usebox{\plotpoint}}
\put(505,458){\usebox{\plotpoint}}
\put(507,459){\usebox{\plotpoint}}
\put(509,460){\rule[-0.350pt]{1.204pt}{0.700pt}}
\put(514,461){\usebox{\plotpoint}}
\put(516,462){\usebox{\plotpoint}}
\put(518,463){\usebox{\plotpoint}}
\put(520,464){\usebox{\plotpoint}}
\put(522,465){\rule[-0.350pt]{0.723pt}{0.700pt}}
\put(525,466){\rule[-0.350pt]{0.964pt}{0.700pt}}
\put(529,467){\usebox{\plotpoint}}
\put(530,468){\usebox{\plotpoint}}
\put(532,469){\rule[-0.350pt]{0.723pt}{0.700pt}}
\put(535,470){\usebox{\plotpoint}}
\put(537,471){\usebox{\plotpoint}}
\put(539,472){\usebox{\plotpoint}}
\put(542,473){\usebox{\plotpoint}}
\put(544,474){\usebox{\plotpoint}}
\put(547,475){\usebox{\plotpoint}}
\put(549,476){\usebox{\plotpoint}}
\put(551,477){\usebox{\plotpoint}}
\put(554,478){\usebox{\plotpoint}}
\put(556,479){\usebox{\plotpoint}}
\put(559,480){\usebox{\plotpoint}}
\put(561,481){\usebox{\plotpoint}}
\put(563,482){\usebox{\plotpoint}}
\put(565,483){\usebox{\plotpoint}}
\put(567,484){\usebox{\plotpoint}}
\put(569,485){\usebox{\plotpoint}}
\put(571,486){\usebox{\plotpoint}}
\put(573,487){\usebox{\plotpoint}}
\put(576,488){\usebox{\plotpoint}}
\put(578,489){\usebox{\plotpoint}}
\put(580,490){\usebox{\plotpoint}}
\put(582,491){\usebox{\plotpoint}}
\put(584,492){\usebox{\plotpoint}}
\put(586,493){\usebox{\plotpoint}}
\put(589,494){\usebox{\plotpoint}}
\put(591,495){\usebox{\plotpoint}}
\put(593,496){\usebox{\plotpoint}}
\put(595,497){\usebox{\plotpoint}}
\put(597,498){\usebox{\plotpoint}}
\put(599,499){\usebox{\plotpoint}}
\put(601,500){\usebox{\plotpoint}}
\put(603,501){\usebox{\plotpoint}}
\put(605,502){\usebox{\plotpoint}}
\put(608,503){\usebox{\plotpoint}}
\put(610,504){\usebox{\plotpoint}}
\put(612,505){\usebox{\plotpoint}}
\put(614,506){\usebox{\plotpoint}}
\put(616,507){\usebox{\plotpoint}}
\put(618,508){\usebox{\plotpoint}}
\put(620,509){\usebox{\plotpoint}}
\put(623,510){\usebox{\plotpoint}}
\put(625,511){\usebox{\plotpoint}}
\put(627,512){\usebox{\plotpoint}}
\put(629,513){\usebox{\plotpoint}}
\put(631,514){\usebox{\plotpoint}}
\put(633,515){\usebox{\plotpoint}}
\put(635,516){\usebox{\plotpoint}}
\put(636,517){\usebox{\plotpoint}}
\put(638,518){\usebox{\plotpoint}}
\put(640,519){\usebox{\plotpoint}}
\put(642,520){\usebox{\plotpoint}}
\put(645,521){\usebox{\plotpoint}}
\put(646,522){\usebox{\plotpoint}}
\put(647,523){\usebox{\plotpoint}}
\put(649,524){\usebox{\plotpoint}}
\put(651,525){\usebox{\plotpoint}}
\put(653,526){\usebox{\plotpoint}}
\put(655,527){\usebox{\plotpoint}}
\put(657,528){\usebox{\plotpoint}}
\put(659,529){\usebox{\plotpoint}}
\put(661,530){\usebox{\plotpoint}}
\put(662,531){\usebox{\plotpoint}}
\put(664,532){\rule[-0.350pt]{0.723pt}{0.700pt}}
\put(667,533){\usebox{\plotpoint}}
\put(668,534){\usebox{\plotpoint}}
\put(670,535){\usebox{\plotpoint}}
\put(671,536){\usebox{\plotpoint}}
\put(673,537){\usebox{\plotpoint}}
\put(675,538){\usebox{\plotpoint}}
\put(676,539){\usebox{\plotpoint}}
\put(678,540){\usebox{\plotpoint}}
\put(680,541){\usebox{\plotpoint}}
\put(682,542){\usebox{\plotpoint}}
\put(683,543){\usebox{\plotpoint}}
\put(685,544){\usebox{\plotpoint}}
\put(687,545){\usebox{\plotpoint}}
\put(688,546){\usebox{\plotpoint}}
\put(690,547){\usebox{\plotpoint}}
\put(692,548){\usebox{\plotpoint}}
\put(694,549){\usebox{\plotpoint}}
\put(695,550){\usebox{\plotpoint}}
\put(697,551){\usebox{\plotpoint}}
\put(699,552){\usebox{\plotpoint}}
\put(700,553){\usebox{\plotpoint}}
\put(702,554){\usebox{\plotpoint}}
\put(704,555){\usebox{\plotpoint}}
\put(705,556){\usebox{\plotpoint}}
\put(707,557){\usebox{\plotpoint}}
\put(709,558){\usebox{\plotpoint}}
\put(711,559){\usebox{\plotpoint}}
\put(712,560){\usebox{\plotpoint}}
\put(714,561){\usebox{\plotpoint}}
\put(715,562){\usebox{\plotpoint}}
\put(717,563){\usebox{\plotpoint}}
\put(719,564){\usebox{\plotpoint}}
\put(720,565){\usebox{\plotpoint}}
\put(722,566){\usebox{\plotpoint}}
\put(724,567){\usebox{\plotpoint}}
\put(725,568){\usebox{\plotpoint}}
\put(727,569){\usebox{\plotpoint}}
\put(728,570){\usebox{\plotpoint}}
\put(730,571){\usebox{\plotpoint}}
\put(731,572){\usebox{\plotpoint}}
\put(733,573){\usebox{\plotpoint}}
\put(735,574){\usebox{\plotpoint}}
\put(736,575){\usebox{\plotpoint}}
\put(738,576){\usebox{\plotpoint}}
\put(739,577){\usebox{\plotpoint}}
\put(741,578){\usebox{\plotpoint}}
\put(742,579){\usebox{\plotpoint}}
\put(744,580){\usebox{\plotpoint}}
\put(745,581){\usebox{\plotpoint}}
\put(746,582){\usebox{\plotpoint}}
\put(748,583){\usebox{\plotpoint}}
\put(749,584){\usebox{\plotpoint}}
\put(750,585){\usebox{\plotpoint}}
\put(752,586){\usebox{\plotpoint}}
\put(753,587){\usebox{\plotpoint}}
\put(755,588){\usebox{\plotpoint}}
\put(757,589){\usebox{\plotpoint}}
\put(759,590){\usebox{\plotpoint}}
\put(760,591){\usebox{\plotpoint}}
\put(761,592){\usebox{\plotpoint}}
\put(762,593){\usebox{\plotpoint}}
\put(763,594){\usebox{\plotpoint}}
\put(765,595){\usebox{\plotpoint}}
\put(766,596){\usebox{\plotpoint}}
\put(767,597){\usebox{\plotpoint}}
\put(768,598){\usebox{\plotpoint}}
\put(770,599){\usebox{\plotpoint}}
\put(771,600){\usebox{\plotpoint}}
\put(773,601){\usebox{\plotpoint}}
\put(774,602){\usebox{\plotpoint}}
\put(776,603){\usebox{\plotpoint}}
\put(777,604){\usebox{\plotpoint}}
\put(778,605){\usebox{\plotpoint}}
\put(780,606){\usebox{\plotpoint}}
\put(781,607){\usebox{\plotpoint}}
\put(783,608){\usebox{\plotpoint}}
\put(785,609){\usebox{\plotpoint}}
\put(786,610){\usebox{\plotpoint}}
\put(787,611){\usebox{\plotpoint}}
\put(789,612){\usebox{\plotpoint}}
\put(790,613){\usebox{\plotpoint}}
\put(791,614){\usebox{\plotpoint}}
\put(792,615){\usebox{\plotpoint}}
\put(793,616){\usebox{\plotpoint}}
\put(794,617){\usebox{\plotpoint}}
\put(796,618){\usebox{\plotpoint}}
\put(797,619){\usebox{\plotpoint}}
\put(798,620){\usebox{\plotpoint}}
\put(799,621){\usebox{\plotpoint}}
\put(801,622){\usebox{\plotpoint}}
\put(803,623){\usebox{\plotpoint}}
\put(804,624){\usebox{\plotpoint}}
\put(805,625){\usebox{\plotpoint}}
\put(806,626){\usebox{\plotpoint}}
\put(807,627){\usebox{\plotpoint}}
\put(808,628){\usebox{\plotpoint}}
\put(809,629){\usebox{\plotpoint}}
\put(810,630){\usebox{\plotpoint}}
\put(812,631){\usebox{\plotpoint}}
\put(813,632){\usebox{\plotpoint}}
\put(814,633){\usebox{\plotpoint}}
\put(815,634){\usebox{\plotpoint}}
\put(816,635){\usebox{\plotpoint}}
\put(818,636){\usebox{\plotpoint}}
\put(819,637){\usebox{\plotpoint}}
\put(820,638){\usebox{\plotpoint}}
\put(821,639){\usebox{\plotpoint}}
\put(822,640){\usebox{\plotpoint}}
\put(824,641){\usebox{\plotpoint}}
\put(825,642){\usebox{\plotpoint}}
\put(826,643){\usebox{\plotpoint}}
\put(827,644){\usebox{\plotpoint}}
\put(829,645){\usebox{\plotpoint}}
\put(830,646){\usebox{\plotpoint}}
\put(831,647){\usebox{\plotpoint}}
\put(832,648){\usebox{\plotpoint}}
\put(833,649){\usebox{\plotpoint}}
\put(834,650){\usebox{\plotpoint}}
\put(835,651){\usebox{\plotpoint}}
\put(836,652){\usebox{\plotpoint}}
\put(838,653){\usebox{\plotpoint}}
\put(839,654){\usebox{\plotpoint}}
\put(840,655){\usebox{\plotpoint}}
\put(841,656){\usebox{\plotpoint}}
\put(842,657){\usebox{\plotpoint}}
\put(843,658){\usebox{\plotpoint}}
\put(844,659){\usebox{\plotpoint}}
\put(846,660){\usebox{\plotpoint}}
\put(847,661){\usebox{\plotpoint}}
\put(848,662){\usebox{\plotpoint}}
\put(849,663){\usebox{\plotpoint}}
\put(850,664){\usebox{\plotpoint}}
\put(851,665){\usebox{\plotpoint}}
\put(852,666){\usebox{\plotpoint}}
\put(853,667){\usebox{\plotpoint}}
\put(854,668){\usebox{\plotpoint}}
\put(855,669){\usebox{\plotpoint}}
\put(856,670){\usebox{\plotpoint}}
\put(857,671){\usebox{\plotpoint}}
\put(859,672){\usebox{\plotpoint}}
\put(860,673){\usebox{\plotpoint}}
\put(861,674){\usebox{\plotpoint}}
\put(862,675){\usebox{\plotpoint}}
\put(863,676){\usebox{\plotpoint}}
\put(864,677){\usebox{\plotpoint}}
\put(865,678){\usebox{\plotpoint}}
\put(866,679){\usebox{\plotpoint}}
\put(867,680){\usebox{\plotpoint}}
\put(868,681){\usebox{\plotpoint}}
\put(869,682){\usebox{\plotpoint}}
\put(870,683){\usebox{\plotpoint}}
\put(871,684){\usebox{\plotpoint}}
\put(872,685){\usebox{\plotpoint}}
\put(873,686){\usebox{\plotpoint}}
\put(874,687){\usebox{\plotpoint}}
\put(875,688){\usebox{\plotpoint}}
\put(876,689){\usebox{\plotpoint}}
\put(877,690){\usebox{\plotpoint}}
\put(878,691){\usebox{\plotpoint}}
\put(879,692){\usebox{\plotpoint}}
\put(880,693){\usebox{\plotpoint}}
\put(881,694){\usebox{\plotpoint}}
\put(882,695){\usebox{\plotpoint}}
\put(883,696){\usebox{\plotpoint}}
\put(884,697){\usebox{\plotpoint}}
\put(885,698){\usebox{\plotpoint}}
\put(886,699){\usebox{\plotpoint}}
\put(887,700){\usebox{\plotpoint}}
\put(888,701){\usebox{\plotpoint}}
\put(889,702){\usebox{\plotpoint}}
\put(890,703){\usebox{\plotpoint}}
\put(891,704){\usebox{\plotpoint}}
\put(892,705){\usebox{\plotpoint}}
\put(893,706){\usebox{\plotpoint}}
\put(894,707){\usebox{\plotpoint}}
\put(895,708){\usebox{\plotpoint}}
\put(896,709){\usebox{\plotpoint}}
\put(897,710){\usebox{\plotpoint}}
\put(898,711){\usebox{\plotpoint}}
\put(899,712){\usebox{\plotpoint}}
\put(900,713){\usebox{\plotpoint}}
\put(901,714){\usebox{\plotpoint}}
\put(902,715){\usebox{\plotpoint}}
\put(903,716){\usebox{\plotpoint}}
\put(904,717){\usebox{\plotpoint}}
\put(905,718){\usebox{\plotpoint}}
\put(906,719){\usebox{\plotpoint}}
\put(907,720){\usebox{\plotpoint}}
\put(908,721){\usebox{\plotpoint}}
\put(909,722){\usebox{\plotpoint}}
\put(910,723){\usebox{\plotpoint}}
\put(911,725){\usebox{\plotpoint}}
\put(912,726){\usebox{\plotpoint}}
\put(913,727){\usebox{\plotpoint}}
\put(914,728){\usebox{\plotpoint}}
\put(915,729){\usebox{\plotpoint}}
\put(916,730){\usebox{\plotpoint}}
\put(917,731){\usebox{\plotpoint}}
\put(918,732){\usebox{\plotpoint}}
\put(919,733){\usebox{\plotpoint}}
\put(920,735){\usebox{\plotpoint}}
\put(921,736){\usebox{\plotpoint}}
\put(922,737){\usebox{\plotpoint}}
\put(923,738){\usebox{\plotpoint}}
\put(924,740){\usebox{\plotpoint}}
\put(925,741){\usebox{\plotpoint}}
\put(926,742){\usebox{\plotpoint}}
\put(927,743){\usebox{\plotpoint}}
\put(928,745){\usebox{\plotpoint}}
\put(929,746){\usebox{\plotpoint}}
\put(930,747){\usebox{\plotpoint}}
\put(931,749){\usebox{\plotpoint}}
\put(931,749){\usebox{\plotpoint}}
\put(932,750){\usebox{\plotpoint}}
\put(933,751){\usebox{\plotpoint}}
\put(934,752){\usebox{\plotpoint}}
\put(935,753){\usebox{\plotpoint}}
\put(936,754){\usebox{\plotpoint}}
\put(937,755){\usebox{\plotpoint}}
\put(938,757){\usebox{\plotpoint}}
\put(939,758){\usebox{\plotpoint}}
\put(940,759){\usebox{\plotpoint}}
\put(941,761){\usebox{\plotpoint}}
\put(264,218){\rule[-0.350pt]{1.975pt}{0.700pt}}
\put(272,219){\rule[-0.350pt]{1.975pt}{0.700pt}}
\put(280,220){\rule[-0.350pt]{1.975pt}{0.700pt}}
\put(288,221){\rule[-0.350pt]{1.975pt}{0.700pt}}
\put(296,222){\rule[-0.350pt]{1.975pt}{0.700pt}}
\put(305,223){\rule[-0.350pt]{1.927pt}{0.700pt}}
\put(313,224){\rule[-0.350pt]{1.927pt}{0.700pt}}
\put(321,225){\rule[-0.350pt]{1.927pt}{0.700pt}}
\put(329,226){\rule[-0.350pt]{1.365pt}{0.700pt}}
\put(334,227){\rule[-0.350pt]{1.365pt}{0.700pt}}
\put(340,228){\rule[-0.350pt]{1.365pt}{0.700pt}}
\put(346,229){\rule[-0.350pt]{1.566pt}{0.700pt}}
\put(352,230){\rule[-0.350pt]{1.566pt}{0.700pt}}
\put(359,231){\rule[-0.350pt]{2.409pt}{0.700pt}}
\put(369,232){\rule[-0.350pt]{1.084pt}{0.700pt}}
\put(373,233){\rule[-0.350pt]{1.084pt}{0.700pt}}
\put(378,234){\rule[-0.350pt]{1.927pt}{0.700pt}}
\put(386,235){\rule[-0.350pt]{1.686pt}{0.700pt}}
\put(393,236){\rule[-0.350pt]{1.445pt}{0.700pt}}
\put(399,237){\rule[-0.350pt]{1.445pt}{0.700pt}}
\put(405,238){\rule[-0.350pt]{1.445pt}{0.700pt}}
\put(411,239){\rule[-0.350pt]{1.445pt}{0.700pt}}
\put(417,240){\rule[-0.350pt]{1.445pt}{0.700pt}}
\put(423,241){\rule[-0.350pt]{1.365pt}{0.700pt}}
\put(428,242){\rule[-0.350pt]{1.365pt}{0.700pt}}
\put(434,243){\rule[-0.350pt]{1.365pt}{0.700pt}}
\put(440,244){\rule[-0.350pt]{1.566pt}{0.700pt}}
\put(446,245){\rule[-0.350pt]{1.566pt}{0.700pt}}
\put(453,246){\rule[-0.350pt]{1.325pt}{0.700pt}}
\put(458,247){\rule[-0.350pt]{1.325pt}{0.700pt}}
\put(464,248){\rule[-0.350pt]{1.084pt}{0.700pt}}
\put(468,249){\rule[-0.350pt]{1.084pt}{0.700pt}}
\put(473,250){\rule[-0.350pt]{1.927pt}{0.700pt}}
\put(481,251){\rule[-0.350pt]{1.686pt}{0.700pt}}
\put(488,252){\rule[-0.350pt]{1.445pt}{0.700pt}}
\put(494,253){\rule[-0.350pt]{1.445pt}{0.700pt}}
\put(500,254){\rule[-0.350pt]{1.204pt}{0.700pt}}
\put(505,255){\rule[-0.350pt]{0.964pt}{0.700pt}}
\put(509,256){\rule[-0.350pt]{1.204pt}{0.700pt}}
\put(514,257){\rule[-0.350pt]{0.964pt}{0.700pt}}
\put(518,258){\rule[-0.350pt]{0.964pt}{0.700pt}}
\put(522,259){\rule[-0.350pt]{0.723pt}{0.700pt}}
\put(525,260){\rule[-0.350pt]{1.686pt}{0.700pt}}
\put(532,261){\rule[-0.350pt]{0.723pt}{0.700pt}}
\put(535,262){\rule[-0.350pt]{1.156pt}{0.700pt}}
\put(539,263){\rule[-0.350pt]{1.156pt}{0.700pt}}
\put(544,264){\rule[-0.350pt]{1.156pt}{0.700pt}}
\put(549,265){\rule[-0.350pt]{1.156pt}{0.700pt}}
\put(554,266){\rule[-0.350pt]{1.156pt}{0.700pt}}
\put(559,267){\rule[-0.350pt]{1.365pt}{0.700pt}}
\put(564,268){\rule[-0.350pt]{1.365pt}{0.700pt}}
\put(570,269){\rule[-0.350pt]{1.365pt}{0.700pt}}
\put(576,270){\rule[-0.350pt]{1.044pt}{0.700pt}}
\put(580,271){\rule[-0.350pt]{1.044pt}{0.700pt}}
\put(584,272){\rule[-0.350pt]{1.044pt}{0.700pt}}
\put(589,273){\rule[-0.350pt]{0.803pt}{0.700pt}}
\put(592,274){\rule[-0.350pt]{0.803pt}{0.700pt}}
\put(595,275){\rule[-0.350pt]{0.803pt}{0.700pt}}
\put(599,276){\rule[-0.350pt]{1.084pt}{0.700pt}}
\put(603,277){\rule[-0.350pt]{1.084pt}{0.700pt}}
\put(608,278){\rule[-0.350pt]{0.964pt}{0.700pt}}
\put(612,279){\rule[-0.350pt]{0.964pt}{0.700pt}}
\put(616,280){\rule[-0.350pt]{0.843pt}{0.700pt}}
\put(619,281){\rule[-0.350pt]{0.843pt}{0.700pt}}
\put(623,282){\rule[-0.350pt]{1.445pt}{0.700pt}}
\put(629,283){\rule[-0.350pt]{0.723pt}{0.700pt}}
\put(632,284){\rule[-0.350pt]{0.723pt}{0.700pt}}
\put(635,285){\rule[-0.350pt]{1.204pt}{0.700pt}}
\put(640,286){\rule[-0.350pt]{1.204pt}{0.700pt}}
\put(645,287){\rule[-0.350pt]{0.964pt}{0.700pt}}
\put(649,288){\rule[-0.350pt]{0.964pt}{0.700pt}}
\put(653,289){\usebox{\plotpoint}}
\put(655,290){\usebox{\plotpoint}}
\put(657,291){\rule[-0.350pt]{1.686pt}{0.700pt}}
\put(664,292){\rule[-0.350pt]{0.723pt}{0.700pt}}
\put(667,293){\rule[-0.350pt]{0.723pt}{0.700pt}}
\put(670,294){\rule[-0.350pt]{0.826pt}{0.700pt}}
\put(673,295){\rule[-0.350pt]{0.826pt}{0.700pt}}
\put(676,296){\rule[-0.350pt]{0.826pt}{0.700pt}}
\put(680,297){\rule[-0.350pt]{0.826pt}{0.700pt}}
\put(683,298){\rule[-0.350pt]{0.826pt}{0.700pt}}
\put(687,299){\rule[-0.350pt]{0.826pt}{0.700pt}}
\put(690,300){\rule[-0.350pt]{0.826pt}{0.700pt}}
\put(694,301){\rule[-0.350pt]{0.819pt}{0.700pt}}
\put(697,302){\rule[-0.350pt]{0.819pt}{0.700pt}}
\put(700,303){\rule[-0.350pt]{0.819pt}{0.700pt}}
\put(704,304){\rule[-0.350pt]{0.819pt}{0.700pt}}
\put(707,305){\rule[-0.350pt]{0.819pt}{0.700pt}}
\put(711,306){\rule[-0.350pt]{0.783pt}{0.700pt}}
\put(714,307){\rule[-0.350pt]{0.783pt}{0.700pt}}
\put(717,308){\rule[-0.350pt]{0.783pt}{0.700pt}}
\put(720,309){\rule[-0.350pt]{0.783pt}{0.700pt}}
\put(724,310){\usebox{\plotpoint}}
\put(726,311){\usebox{\plotpoint}}
\put(729,312){\usebox{\plotpoint}}
\put(732,313){\usebox{\plotpoint}}
\put(735,314){\rule[-0.350pt]{0.723pt}{0.700pt}}
\put(738,315){\rule[-0.350pt]{0.723pt}{0.700pt}}
\put(741,316){\rule[-0.350pt]{0.723pt}{0.700pt}}
\put(744,317){\rule[-0.350pt]{0.964pt}{0.700pt}}
\put(748,318){\rule[-0.350pt]{0.964pt}{0.700pt}}
\put(752,319){\usebox{\plotpoint}}
\put(754,320){\usebox{\plotpoint}}
\put(756,321){\usebox{\plotpoint}}
\put(759,322){\rule[-0.350pt]{0.723pt}{0.700pt}}
\put(762,323){\rule[-0.350pt]{0.723pt}{0.700pt}}
\put(765,324){\usebox{\plotpoint}}
\put(767,325){\usebox{\plotpoint}}
\put(770,326){\rule[-0.350pt]{0.723pt}{0.700pt}}
\put(773,327){\rule[-0.350pt]{0.723pt}{0.700pt}}
\put(776,328){\usebox{\plotpoint}}
\put(778,329){\usebox{\plotpoint}}
\put(780,330){\rule[-0.350pt]{1.204pt}{0.700pt}}
\put(785,331){\usebox{\plotpoint}}
\put(787,332){\usebox{\plotpoint}}
\put(789,333){\rule[-0.350pt]{0.723pt}{0.700pt}}
\put(792,334){\usebox{\plotpoint}}
\put(794,335){\usebox{\plotpoint}}
\put(796,336){\rule[-0.350pt]{0.723pt}{0.700pt}}
\put(799,337){\rule[-0.350pt]{0.964pt}{0.700pt}}
\put(803,338){\rule[-0.350pt]{0.723pt}{0.700pt}}
\put(806,339){\usebox{\plotpoint}}
\put(808,340){\usebox{\plotpoint}}
\put(810,341){\usebox{\plotpoint}}
\put(812,342){\usebox{\plotpoint}}
\put(815,343){\usebox{\plotpoint}}
\put(817,344){\usebox{\plotpoint}}
\put(819,345){\usebox{\plotpoint}}
\put(822,346){\usebox{\plotpoint}}
\put(824,347){\usebox{\plotpoint}}
\put(826,348){\usebox{\plotpoint}}
\put(829,349){\usebox{\plotpoint}}
\put(831,350){\usebox{\plotpoint}}
\put(833,351){\usebox{\plotpoint}}
\put(835,352){\usebox{\plotpoint}}
\put(837,353){\usebox{\plotpoint}}
\put(839,354){\usebox{\plotpoint}}
\put(841,355){\usebox{\plotpoint}}
\put(843,356){\usebox{\plotpoint}}
\put(846,357){\usebox{\plotpoint}}
\put(848,358){\usebox{\plotpoint}}
\put(850,359){\usebox{\plotpoint}}
\put(852,360){\usebox{\plotpoint}}
\put(854,361){\usebox{\plotpoint}}
\put(856,362){\usebox{\plotpoint}}
\put(859,363){\usebox{\plotpoint}}
\put(861,364){\usebox{\plotpoint}}
\put(863,365){\usebox{\plotpoint}}
\put(865,366){\usebox{\plotpoint}}
\put(867,367){\usebox{\plotpoint}}
\put(870,368){\usebox{\plotpoint}}
\put(872,369){\usebox{\plotpoint}}
\put(874,370){\usebox{\plotpoint}}
\put(876,371){\usebox{\plotpoint}}
\put(879,372){\usebox{\plotpoint}}
\put(881,373){\usebox{\plotpoint}}
\put(883,374){\usebox{\plotpoint}}
\put(885,375){\usebox{\plotpoint}}
\put(887,376){\usebox{\plotpoint}}
\put(888,377){\usebox{\plotpoint}}
\put(890,378){\usebox{\plotpoint}}
\put(892,379){\usebox{\plotpoint}}
\put(894,380){\usebox{\plotpoint}}
\put(896,381){\usebox{\plotpoint}}
\put(898,382){\usebox{\plotpoint}}
\put(900,383){\usebox{\plotpoint}}
\put(902,384){\usebox{\plotpoint}}
\put(904,385){\usebox{\plotpoint}}
\put(906,386){\usebox{\plotpoint}}
\put(907,387){\usebox{\plotpoint}}
\put(909,388){\usebox{\plotpoint}}
\put(911,389){\usebox{\plotpoint}}
\put(912,390){\usebox{\plotpoint}}
\put(914,391){\usebox{\plotpoint}}
\put(916,392){\usebox{\plotpoint}}
\put(918,393){\usebox{\plotpoint}}
\put(920,394){\usebox{\plotpoint}}
\put(921,395){\usebox{\plotpoint}}
\put(922,396){\usebox{\plotpoint}}
\put(924,397){\usebox{\plotpoint}}
\put(926,398){\usebox{\plotpoint}}
\put(928,399){\usebox{\plotpoint}}
\put(929,400){\usebox{\plotpoint}}
\put(931,401){\usebox{\plotpoint}}
\put(933,402){\usebox{\plotpoint}}
\put(935,403){\usebox{\plotpoint}}
\put(936,404){\usebox{\plotpoint}}
\put(938,405){\usebox{\plotpoint}}
\put(939,406){\usebox{\plotpoint}}
\end{picture}
\caption{The temperatures $T_{\cal R}$ as a function of Hall defect
	 ${\cal R}$ for (a) $\nu =1 $ and (b) $\nu =2 $ for a
	 typical value of $\rho /m^\ast =20$ cm$^{-1}$.}
\end{figure}
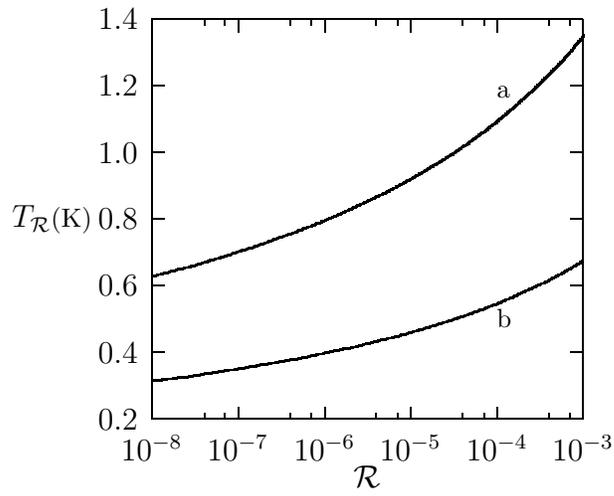

\end{document}